# OpenMM 8: Molecular Dynamics Simulation with Machine Learning Potentials

Peter Eastman[1]*, Raimondas Galvelis[2,3], Raúl P. Peláez[3], Charlles R. A. Abreu[4,5], Stephen E. Farr[6], Emilio Gallicchio[7,8], Anton Gorenko[9], Michael M. Henry[10], Frank Hu[1], Jing Huang[11], Andreas Krämer[12], Julien Michel[6], Joshua A. Mitchell[13], Vijay S. Pande[14,15], João PGLM Rodrigues[15], Jaime Rodriguez-Guerra[16], Andrew C. Simmonett[17], Sukrit Singh[10], Jason Swails[18], Philip Turner[19], Yuanqing Wang[20], Ivy Zhang[10,21], John D. Chodera[10], Gianni De Fabritiis[2,3,22], Thomas E. Markland[1]

[1]Department of Chemistry, Stanford University, Stanford, CA 94305, USA
[2]Acellera Labs, C Dr Trueta 183, 08005, Barcelona, Spain
[3]Computational Science Laboratory, Universitat Pompeu Fabra, Barcelona Biomedical Research Park (PRBB), C Dr. Aiguader 88, 08003, Barcelona, Spain
[4]Chemical Engineering Department, School of Chemistry, Federal University of Rio de Janeiro, Rio de Janeiro 68542, Brazil
[5]Redesign Science Inc., 180 Varick St., New York, NY 10014, USA
[6]EaStCHEM School of Chemistry, University of Edinburgh, EH9 3FJ, United Kingdom
[7]Department of Chemistry and Biochemistry, Brooklyn College of the City University of New York, NY, USA
[8]Ph.D. Program in Chemistry and Ph.D. Program in Biochemistry, The Graduate Center of the City University of New York, New York, NY, USA
[9]Stream HPC, Koningin Wilhelminaplein 1 - 40601, 1062 HG Amsterdam, Netherlands
[10]Computational and Systems Biology Program, Sloan Kettering Institute, Memorial Sloan Kettering Cancer Center, New York NY 10065, USA
[11]Key Laboratory of Structural Biology of Zhejiang Province, School of Life Sciences, Westlake University, 18 Shilongshan Road, Hangzhou 310024, Zhejiang, China
[12]Department of Mathematics and Computer Science, Freie Universität Berlin, Arnimallee 12, 14195 Berlin, Germany
[13]The Open Force Field Initiative, Open Molecular Software Foundation, Davis, CA 95616, USA
[14]Andreessen Horowitz, 2865 Sand Hill Rd, Menlo Park, CA 94025, USA
[15]Department of Structural Biology, Stanford University, Stanford, CA 94305, USA
[16]Charité Universitätsmedizin Berlin In silico Toxicology and Structural Bioinformatics, Virchowweg 6, 10117 Berlin, Germany
[17]Laboratory of Computational Biology, National Heart, Lung and Blood Institute, National Institutes of Health, Bethesda, MD 20892, USA
[18]Entos Inc., 9310 Athena Circle, La Jolla, CA 92037, USA
[19]College of Engineering, Virginia Polytechnic Institute and State University, Blacksburg, VA 24061, USA
[20]Simons Center for Computational Physical Chemistry and Center for Data Science, New York University, 24 Waverly Place, New York, NY 10004, USA
[21]Tri-Institutional PhD Program in Computational Biology and Medicine, Weill Cornell Medical College, Cornell University, New York, NY 10065, USA
[22]ICREA, Passeig Lluis Companys 23, 08010, Barcelona, Spain

*Corresponding author (peastman@stanford.edu)

# Abstract

Machine learning plays an important and growing role in molecular simulation. The newest version of the OpenMM molecular dynamics toolkit introduces new features to support the use of machine learning potentials. Arbitrary PyTorch models can be added to a simulation and used to compute forces and energy. A higher-level interface allows users to easily model their molecules of interest with general purpose, pretrained potential functions. A collection of optimized CUDA kernels and custom PyTorch operations greatly improves the speed of simulations. We demonstrate these features on simulations of cyclin-dependent kinase 8 (CDK8) and the green fluorescent protein (GFP) chromophore in water. Taken together, these features make it practical to use machine learning to improve the accuracy of simulations at only a modest increase in cost.

# Introduction

In recent years, much work in the field of molecular simulation has focused on ways to produce more accurate results at lower cost. This includes newer force fields that use better functional forms or better parametrization to improve accuracy without significantly increasing cost. It also includes semi-empirical and machine learning methods that try to approach the accuracy of high-level quantum chemistry (QC) methods at a cost that is intermediate between QC and classical force fields. Another important aspect is the use of newer sampling methods to reduce the amount of simulation needed to obtain converged results for thermodynamic quantities.

In this paper, we describe the newly released version 8 of OpenMM, a popular package for molecular simulation that provides excellent performance and high flexibility.[1] It contains new features to better support these methods, with a particular emphasis on machine learning. They make it a powerful tool both for methodological research and for production simulations.

Machine learning potentials (MLPs) have increasingly emerged as a major area of research within the field of molecular simulation. They offer a powerful intermediate between accurate but slow quantum chemistry methods on the one hand, and fast but less accurate force fields on the other. Unlike force fields, which use simple functions chosen by the designer to compute interactions, MLPs use highly flexible models that can learn very complicated functions, most often implemented as neural networks. The models are trained on molecular energies, forces, and other properties computed with a high-level QC method. The trained models can approach the accuracy of the method used to generate the training data, while being orders of magnitude faster.[2]

The past few years have seen an explosion of new MLP architectures designed to improve their accuracy, speed, transferability, and data efficiency. Many of these architectures try to mirror the symmetries obeyed by the physical system, particularly equivariance under translations and rotations.[3–7] Others try to improve transferability by directly incorporating a limited amount of physics into the model, such as explicit terms for Coulomb interactions, dispersion, and nuclear repulsion.[8–11] Still others take this approach even further, beginning with a semi-empirical quantum chemistry method and training a machine learning model to correct for its flaws and improve its accuracy.[12–14] These approaches provide different tradeoffs between speed, accuracy, and range of applicability, with new architectures continuing to be introduced.

Another important area of research is ways of combining MLPs with classical force fields. Although they are much faster than high-level quantum chemistry algorithms, they are still orders of magnitude slower than most force fields. This often makes them impractical for simulating macromolecules or other large systems. Just as QM/MM simulations use quantum mechanics for only a small piece of a system while simulating the rest with classical mechanics, ML/MM simulations can simulate part with machine learning and the rest with a classical force field. This can lead to large improvements in accuracy with only a modest increase in computational cost.[15,16]

We have recently released version 8 of the OpenMM simulation package, with a focus on supporting machine learning potentials. OpenMM is a widely used piece of software for molecular simulation, especially biological macromolecules, and one of the fastest growing engines for atomistic molecular simulations[17]. Its architecture and features provide a unique combination of speed, flexibility, and extensibility. Some of the more notable aspects include its excellent performance when running on GPUs; the ability to implement entirely new algorithms and simulation protocols through Python scripting; and the ability to define entirely new functional forms for interactions, which are transparently compiled to machine code and executed with no loss of performance on the GPU or CPU.

OpenMM 8 extends this flexibility to MLPs. The new features described below aim to achieve two goals. First, they provide a powerful environment for researchers developing new MLPs or new simulation protocols. Arbitrary PyTorch models can be easily added to a simulation and used to compute forces and energies. Second, they bridge the gap to production simulations, allowing end users to select from a set of validated general-purpose MLPs and run simulations with them. With these new features, setting up and running a ML or ML/MM simulation is no more difficult than using a conventional force field.

# Methods

## PyTorch Model Support

The foundation of MLP support is the OpenMM-Torch module. It provides the TorchForce class, which allows arbitrary PyTorch[18] models to be embedded in a simulation and used to calculate forces and energies. PyTorch has emerged as the standard machine learning framework within the simulation community, and nearly all MLP models published in recent years have been implemented with it. By supporting it, we allow existing models to easily interface with OpenMM. It also provides an ideal environment for methodological research, since model developers can continue using the same framework they are already accustomed to. Interfaces to TensorFlow and TensorRT are also available, but because the community has largely standardized on PyTorch, they are not discussed further here.

To create a TorchForce, one simply provides a model that takes atomic positions as input and computes potential energy. Forces can be automatically calculated through backpropagation, or alternatively, the model can return them explicitly. All features of PyTorch are supported. The only restriction on the model is that it must be possible to compile it to TorchScript. The following code is a minimalist example that illustrates how to apply a potential function computed with PyTorch, in this case a harmonic potential attracting every particle to the origin.

```
class ForceModule(torch.nn.Module):
    def forward(self, positions):
        return torch.sum(positions**2)

module = torch.jit.script(ForceModule())
force = TorchForce(module)
```

## OpenMM-ML

The next level in the stack is the OpenMM-ML module, which provides a convenient mechanism for running simulations with pretrained potentials. It makes MLPs as simple to use as conventional force fields. For example, given a Topology object (a description of the atoms and chemical bonds making up a system to be simulated), the following code prepares a simulation of it with the ANI-2x[19] potential function.

```
potential = MLPotential('ani2x')
system = potential.createSystem(topology)
```

Mixed simulations, in which a system is modeled partly with a conventional force field and partly with a MLP, are just as easy to prepare. One simply provides a version that is modeled entirely with the force field, and a list of which atoms make up the subset to model with ML. All details of setting up the mixed system, including handling the interaction of the two regions through mechanical embedding, are handled automatically.

```
forcefield = ForceField('amber14-all.xml', 'amber14/tip3pfb.xml')
mm_system = forcefield.createSystem(topology)
potential = MLPotential('ani2x')
ml_system = potential.createMixedSystem(topology, mm_system, ml_atoms)
```

At present, the only supported MLPs are ANI-1ccx[20] and ANI-2x[19]. These potentials are fast and have reasonably good accuracy, but their range of applicability is limited. They support only a limited set of elements (seven for ANI-2x, four for ANI-1ccx), and do not support charged molecules. We intend that the list of potentials will grow with time as new pretrained MLPs are developed and made available to the community.

To facilitate the development of new models, we have created the SPICE (Small-Molecule/Protein Interaction Chemical Energies) dataset[21], a large collection of molecular forces and energies calculated at the highly accurate ωB97M-D3BJ/def2-TZVPPD[22] level of theory. It incorporates a wide selection of molecules relevant to simulating drug-like small molecules interacting with proteins. It includes 15 elements, both charged and neutral molecules, both low and high energy conformations, and a wide assortment of covalent and non-covalent interactions. We hope that with time it will lead to the creation of many new pretrained MLPs that can be made available for use by the simulation community.

## Optimization Through NNPOps

For MLPs to be useful, they must be fast. For this reason, we created the NNPOps module with the goal of accelerating machine learning models. It contains optimized CUDA kernels and PyTorch code for bottleneck operations in important MLPs. Examples include computing the atom-centered symmetry functions used in the ANI family of potentials, building neighbor lists, and computing electrostatic interactions with the Particle Mesh Ewald method. Benchmarks demonstrating the speedup from NNPOps are in the next section.

## Other Features

While OpenMM 8 includes important new features for machine learning, numerous other new features have been added in the years since version 7 was released in 2016. We highlight here a few of the more significant ones.

**Enhanced Sampling Methods**

Enhanced sampling algorithms allow a simulation to explore conformation space more efficiently and produce converged results in less time. A number of methods are available for use with OpenMM.

Metadynamics[23] is supported through two different mechanisms. One is based on the popular PLUMED library[24], which offers a wide selection of useful collective variables (CVs) along which to accelerate exploration. The other is a native implementation that uses a novel mechanism to define CVs. Any energy term that can be defined with OpenMM's force classes can be used as a CV. Given OpenMM's flexible custom forces supporting arbitrary user-defined functional forms, this is a very powerful mechanism.

To illustrate the flexibility of this approach, we present a few examples of how common CVs can be defined. One common choice is a weighted sum of distances between particular pairs of atoms. This is easily defined with OpenMM's CustomBondForce.

```
cv = CustomBondForce('weight*r')
cv.addPerBondParameter('weight')
for atom1, atom2, weight in pairs:
    cv.addBond(atom1, atom2, [weight])
```

In other cases a weighted sum of dihedral angles could be more appropriate. This is implemented just as easily with CustomTorsionForce.

```
cv = CustomTorsionForce('weight*theta')
cv.addPerTorsionParameter('weight')
for atom1, atom2, atom3, atom4, weight in dihedrals:
    cv.addTorsion(atom1, atom2, atom3, atom4, [weight])
```

More complicated functions of distances, angles, and dihedrals can be defined just as easily, as can terms that depend on the centroids of groups of atoms, root mean squared deviations (RMSD) between sets of atoms, and many other functions. One can even use the TorchForce class described above to define CVs through PyTorch code.

For situations when specific CVs are not known in advance, an implementation of simulated tempering[25] has been added. It accelerates transitions across barriers by allowing a simulation to temporarily increase its temperature. It joins other methods that were already available in OpenMM and its associated packages, including aMD[26], replica exchange[27], and Replica Exchange with Solute Tempering (REST)[28].

Another new algorithm is the Alchemical Transfer Method[29] (ATM), which provides a convenient and efficient method of computing free energy differences. Interaction forces are computed before and after performing a coordinate transformation, such as translating a molecule between the binding pocket and bulk solvent. The simulation evolves based on a weighted average of the two. This permits efficient sampling of alchemical pathways used to compute absolute or relative free energies. Unlike many other alchemical methods, it does not require soft-core interactions and can be applied to arbitrary potential functions, including ones defined by MLPs.

**New Force Fields**

Classical force fields continue to advance, improving both in accuracy and range of applicability. Many new force fields are now supported. This includes newer versions of the Amber[30] and CHARMM[31] point charge force fields and the polarizable AMOEBA[32] and CHARMM Drude[33] force fields. The GLYCAM[34] force field has been added for simulating carbohydrates. For simulating arbitrary small organic molecules, the OpenFF[35] and GAFF[36] generic force fields are now available.

**New Integrators**

OpenMM has added support for the LF-Middle integration algorithm[37,38]. This is an improved discretization of the Langevin equation that significantly reduces error compared to previously used discretizations. In many cases, it allows doubling the step size with no loss in accuracy, leading to a huge increase in performance.

Other new integrators have also been added, including a Nosé-Hoover integrator[39], a dual-temperature version of the Nosé-Hoover integrator for simulating Drude particles[40], and a multiple time step Langevin integrator. In addition, many features have been added to the CustomIntegrator class, which allows users to define entirely new integration algorithms. These features greatly increase the range of methods that can be implemented with it.

**Improved Hardware Support**

OpenMM has grown in its ability to take advantage of a range of hardware. ARM and PowerPC processors are now supported. A new backend based on the HIP framework provides greatly improved performance on AMD GPUs (see Table 1). The ability to efficiently parallelize a simulation across multiple GPUs has improved (see Table 2). The speed of running simulations on CPUs has also greatly improved.

| Molecule | Atoms | OpenCL (ns/day) | HIP (ns/day) |
| --- | --- | --- | --- |
| DHFR | 23,558 | 417 | 1031 |
| ApoA1 | 92,224 | 192 | 393 |
| Cellulose | 408,609 | 42 | 91 |

Table 1. The speed of the OpenCL and HIP platforms simulating three benchmark systems on an AMD V620 GPU. All benchmarks used Langevin dynamics with a 4 fs time step, constraining the lengths of bonds involving hydrogen.

| GPUs | Speed (ns/day) |
| --- | --- |
| 1 | 32.3 |
| 2 | 51.5 |
| 3 | 61.6 |
| 4 | 70.9 |

Table 2. The speed of simulating Satellite Tobacco Mosaic Virus (1,067,095 atoms) on one to four NVIDIA A100 GPUs connected by NVLink. All benchmarks use the CUDA platform, Langevin dynamics, a 4 fs time step, and constraints on the lengths of bonds involving hydrogen.

# Results and Discussion

## Performance of MLPs

To demonstrate the effect of the optimizations in NNPOps, we simulated cyclin-dependent kinase 8 (CDK8) bound to a 53 atom inhibitor (lig_30 from the protein-ligand-benchmark set[41], shown in Figure 1). CDK8 has 632 residues and 10,456 atoms. We modeled it with the Amber14 force field and solvated it with 41,194 TIP3P-FB water molecules, giving a total of 134,099 atoms.

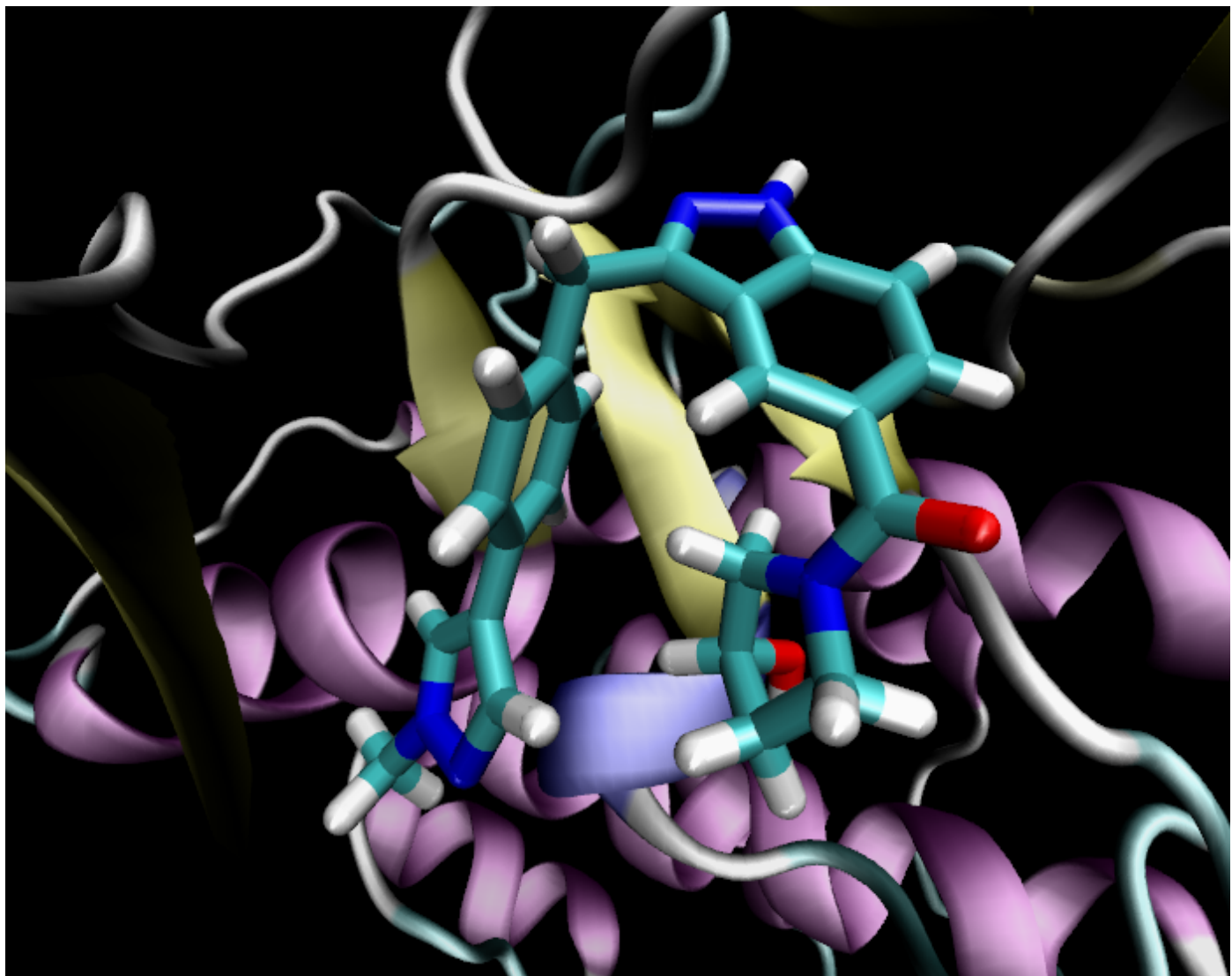

Figure 1: The 53 atom inhibitor bound to CDK8.

We compared the simulation speed when modeling the inhibitor either with the classical OpenFF 2.0.0 force field[35] or with the ANI-2x machine learning potential. Interactions between the inhibitor and the rest of the system were computed with the Amber14 force field in all cases. We further compared the speed of ANI-2x using either the standard implementation from the TorchANI library or the optimized version from NNPOps.

ANI-2x is an ensemble of eight models that were each trained independently. The average of the models is slightly more accurate than any one of the models on its own. When speed is important, it is often acceptable to use just one of the eight models. This produces a large improvement in speed in exchange for a small loss in accuracy. We benchmarked this option as well.

All of the simulations described above were run with a Langevin integrator and a 2 fs time step. In classical simulations, it is common to constrain the lengths of bonds involving hydrogen and to increase the mass of hydrogen atoms through mass repartitioning. This combination allows integration to be stable with a 4 fs time step. In principle, the same optimization could be used for the MLP, but it would require careful selection of the constrained bond lengths and a thorough validation of the results. Since this optimization is not yet common practice in ML/MM simulations, we chose to omit constraints within the ligand and limit the time step to 2 fs. For the classical force field, we measured the speed with both 2 fs and 4 fs time steps.

Each simulation was integrated for 10,000 steps, and the speed in ns/day was calculated based on the elapsed time. These very short simulations are intended only to benchmark speed. Our goal in this section is not to validate the accuracy of the ANI-2x potential, which has been studied elsewhere[19], only to measure the effect of our optimizations.

All simulations were run on a NVIDIA RTX 4080 GPU. The results are shown in Table 3.

| Potential | Implementation | Step Size (fs) | Speed (ns/day) |
| --- | --- | --- | --- |
| ANI-2x (ensemble) | TorchANI | 2 | 9.8 |
| ANI-2x (ensemble) | NNPOps | 2 | 56 |
| ANI-2x (single) | NNPOps | 2 | 105 |
| OpenFF 2.0.0 | | 2 | 281 |
| OpenFF 2.0.0 | | 4 | 531 |

Table 3. The speed of simulating a 53 atom inhibitor bound to CDK8.

NNPOps is faster than TorchANI by a factor of 5.7, enormously increasing the practicality of using the potential for real simulations. Using a single model instead of the full ensemble improves the speed by another factor of 1.9. Combining these two optimizations, modeling the ligand with ANI-2x is only 2.7 times slower than a fully classical simulation with the same step size. This is a very reasonable cost, given that it has been shown to significantly improve the accuracy in calculated binding free energies. Using a larger step size in the classical simulation increases the ratio to 5.1 times slower. In principle, the larger step size could also be used for ML/MM simulations, but one would need to evaluate the effects of constraining bond lengths on the results.

Viewed another way, these results illustrate how expensive current MLPs still are compared to classical force fields; simulating 53 atoms with ANI-2x is more expensive than simulating the other 134,046 atoms with Amber14. On the other hand, it is orders of magnitude faster than conventional QC methods with similar accuracy. Simulating core pieces of a system with MLPs is now entirely practical, and can yield large improvements to accuracy.

## Example: MLP for Anionic Green Fluorescent Protein Chromophore in Water

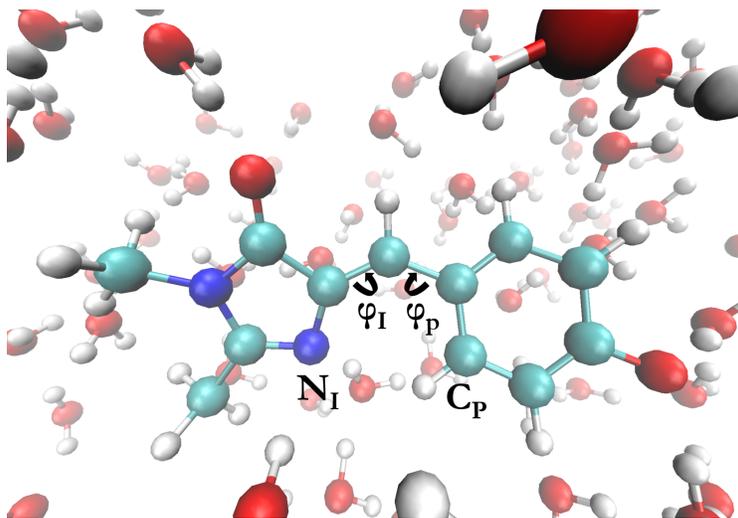

Figure 2: Anionic GFP chromophore in water. The dihedral $\phi_P$ is defined between the $C_P$ atom and the carbons of the methine bridge and the dihedral $\phi_I$ is defined between the $N_I$ atom and the carbons of the methine bridge.

As a demonstration of the TorchForce class and OpenMM's MLP integration capabilities, we trained an MLP to simulate anionic GFP chromophore (*p*-hydroxybenzylidene-2,3-dimethylimidazolinone, HBDI-)[42] in water. GFP is a widely used genetically-encoded fluorescent tag in biological systems that features a covalently attached chromophore formed by an autocatalytic reaction following folding[43]. The chromophore's fluorescent properties are strongly influenced by its environment. For that reason, it is useful to study its behavior in a variety of environments, including water and other solvents. The system consists of the anionic GFP chromophore with 167 solvating waters for a total of 528 atoms in a periodic orthorhombic cell. A picture of the GFP chromophore and the surrounding water molecules can be seen in Figure 2.

The training data was obtained from *ab initio* molecular dynamics (AIMD) trajectories of the GFP chromophore in water (75008 configurations from 10 separate trajectories) supplemented with 7500 configurations obtained from *ab initio* path integral molecular dynamics (AI-PIMD) trajectories, which incorporate nuclear quantum effects in the sampling. In these trajectories, obtained from our previous work[44], forces were evaluated with CP2K[45] using DFT with the revPBE exchange-correlation functional[46,47] and D3 dispersion corrections[48] added. All trajectories used multiple time step (MTS) integration of the r-RESPA form[49], with an outer time step of 2.0 fs and an inner time step of 0.5 fs. The MTS reference forces were evaluated at a SCC-DFTB3 level of theory[50]. The configurations used for training were taken every 2.0 fs and correspond to 15 ps of AI-PIMD trajectory and ~150 ps of classical AIMD trajectories.

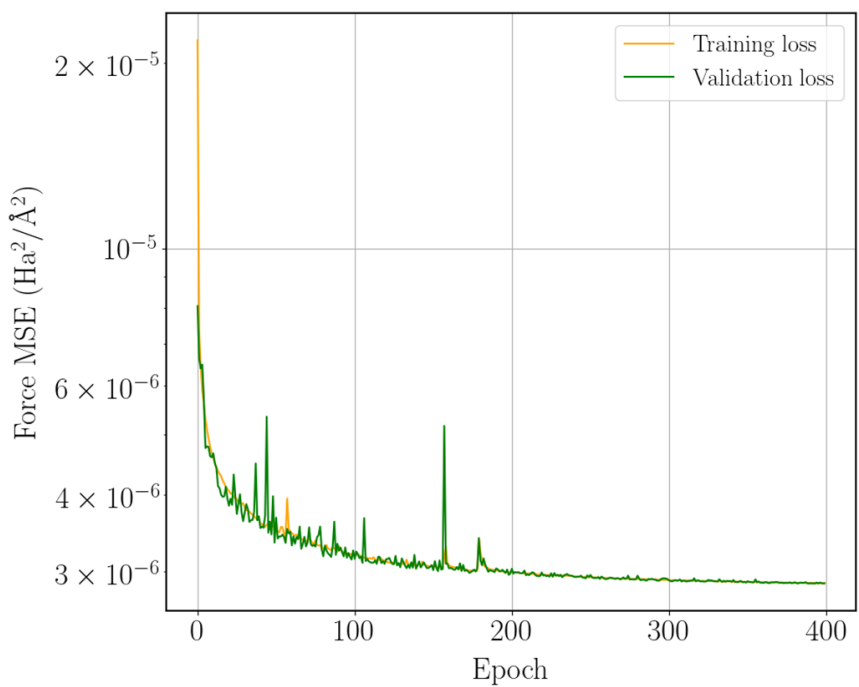

Figure 3: Equivariant transformer training and validation loss.

The machine learning architecture that we employ is an equivariant transformer neural network using the TorchMD-NET package[4]. The training data was partitioned randomly into an 80-10-10 split for the training, validation, and test sets respectively, and the model was trained for 400 epochs to all the forces in each configuration. Our model used 2 attention heads, 1 layer, 32 radial basis functions, and an embedding dimension of 64. This resulted in a test set RMSE on forces of 45.7 meV/A. Learning curves for the model are shown in Figure 3.

The trained model was compiled into a TorchScript module and used the TorchForce class to run dynamics in OpenMM 8.0. Using the LangevinMiddleIntegrator, we ran NVT dynamics with a temperature of 300 K, 0.5 fs time step, and a 1.0 ps$^{-1}$ friction coefficient. The final trajectory used for analysis was 1 ns in length, with frames saved every 2.0 fs. Of this trajectory, we discarded the first 15 ps for equilibration.

For comparing our MLP-generated trajectory to the AIMD trajectories, we analyzed the oxygen-oxygen radial distribution functions (RDFs) around the two oxygen atoms on the GFP chromophore as well as the distribution of the $\phi_I$ and $\phi_P$ dihedral angles as defined in Figure 2. We chose these two observables because (1) the two oxygen atoms on the chromophore are those that interact the most strongly with the solvent water molecules; and (2) the dihedrals $\phi_I$ and $\phi_P$ are the angles that undergo torsional rotation upon photoexcitation of the chromophore[51].

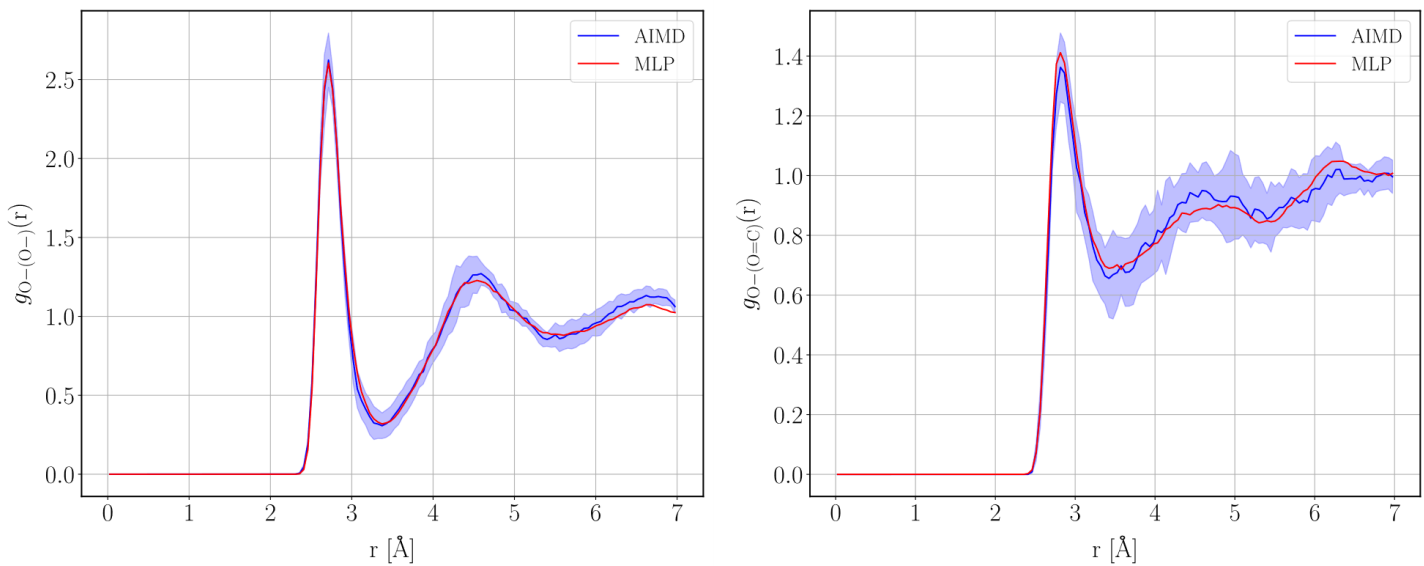

Figure 4: Anionic oxygen-oxygen RDF (left) and carbonyl oxygen-oxygen RDF (right) generated from the MLP and AIMD trajectories. The shaded region represents +/- 1 standard deviation across the 10 AIMD trajectories.

Figure 4 shows the comparison of the RDFs obtained using the MLP and the AIMD where the error bars on the AIMD RDFs are +/- 1 standard deviation across the 10 AIMD trajectories. The MLP and AIMD are in good agreement and within the error bars of the AIMD trajectories. The distribution of dihedral angles $\phi_I$ and $\phi_P$ can be seen in Figure 5 and are in similarly good agreement within the error bars.

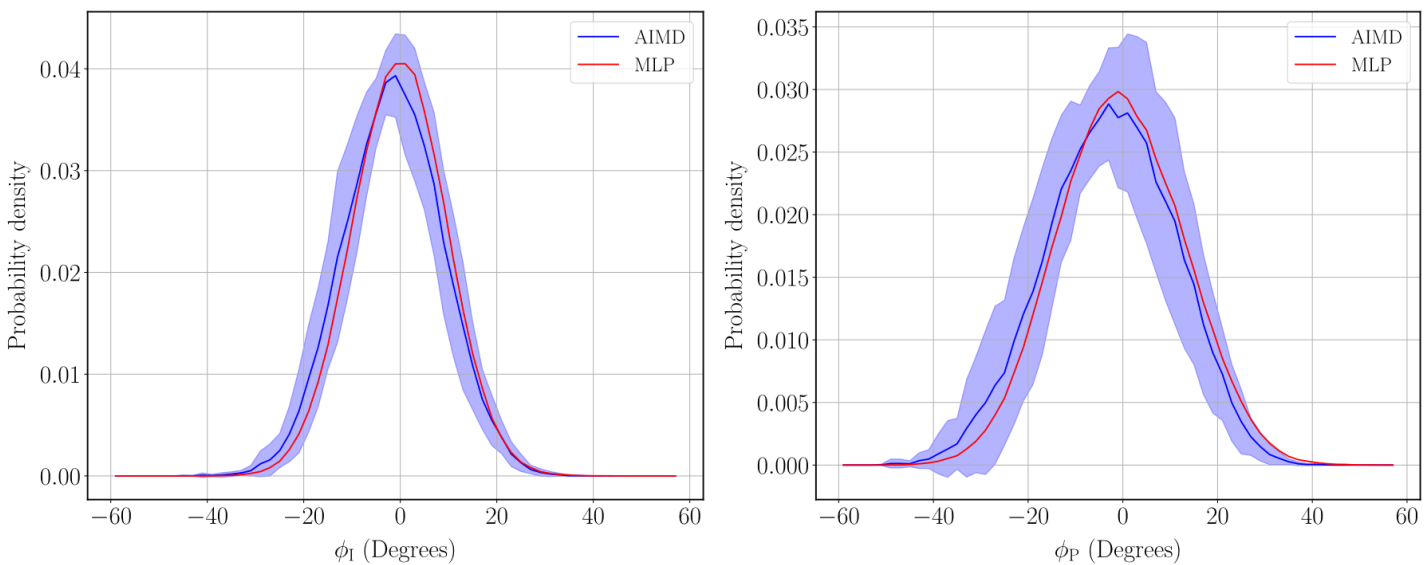

Figure 5: Distribution of dihedral angles for $\phi_I$ and $\phi_P$. The shaded region represents +/- 1 standard deviation across the 10 AIMD trajectories.

The advantages of the MLP are evident from the timings. On two AMD EPYC 7763 64 core processors (128 cores total), the AIMD calculation of the force takes 13.2 seconds, whereas the MLP evaluated on a Nvidia RTX 4080 GPU takes only 8.7 ms for each force evaluation, a speed-up of 1517 times.

# Conclusions

OpenMM 8 introduces new features to support the use of machine learning potentials in molecular simulations. These features are best viewed as a foundation on which to build future tools. MLPs are still an immature field; few pretrained, general-purpose potentials exist, and those that do exist are limited in their range of applicability. More often, users must train their own special purpose potentials, as we demonstrate here for the GFP chromophore. The tools in OpenMM provide a powerful environment for researchers developing new pretrained potentials. Those future potentials can then be made easily available to users who wish to run simulations with them.

The tools for running hybrid ML/MM simulations should likewise be viewed as a starting point for future work. For example, an important feature will be to allow the ML and MM regions to interact through electrostatic embedding, rather than the simpler but less accurate mechanical embedding. This will require ML models that can predict atomic partial charges as well as energy.

With time we hope to provide a diverse set of potentials to meet many different needs. To an even greater extent than classical force fields, MLPs can offer a huge range of tradeoffs between speed, accuracy, and range of applicability. Developing a corresponding range of potentials is an important project for the community.

Availability

OpenMM is available from https://openmm.org. Development is conducted through the Github community at https://github.com/openmm/openmm.

# Acknowledgements

Research reported in this publication was supported by the National Institute of General Medical Sciences (NIGMS) of the National Institutes of Health under award number GM140090 (PE, TEM, JDC, GDF). The content is solely the responsibility of the authors and does not necessarily represent the official views of the National Institutes of Health. PE and TEM acknowledge support from the Chan Zuckerberg Initiative's Essential Open Source Software for Science program grant EOSS2-0000000172. JDC acknowledges support from NIH grant P30 CA008748, NIH grant R01 GM140090, and the Sloan Kettering Institute. SF and JM acknowledge support from the Engineering and Physical Sciences Research Council under grant award EP/W030276/1. GDF acknowledges support from the project PID2020-116564GB-I00 that has been funded by MCIN/AEI/10.13039/501100011033. I.Z. acknowledges support from Vir Biotechnology, Inc., a Molecular Sciences Software Institute Seed Fellowship, and the Tri-Institutional Program in Computational Biology and Medicine. ACS acknowledges support from the intramural research program of the National Heart, Lung and Blood Institute. JPR acknowledges support from NIH NIGMS grant R35GM122543. EG acknowledges support from the NSF CAREER award 1750511. YW acknowledges support from the Schmidt Science Fellowship, in partnership with the Rhodes Trust, and the Simons Center for Computational Physical Chemistry at New York University. SS is a Damon Runyon Quantitative Biology Fellow supported by the Damon Runyon Cancer Research Foundation (DRQ-14-22).

# Disclosures

JDC is a current member of the Scientific Advisory Board of OpenEye Scientific Software, Redesign Science, Ventus Therapeutics, and Interline Therapeutics, and has equity interests in Interline Therapeutics. The Chodera laboratory receives or has received funding from multiple sources, including the National Institutes of Health, the National

Science Foundation, the Parker Institute for Cancer Immunotherapy, Relay Therapeutics, Entasis Therapeutics, Silicon Therapeutics, EMD Serono (Merck KGaA), AstraZeneca, Vir Biotechnology, Bayer, XtalPi, Interline Therapeutics, the Molecular Sciences Software Institute, the Starr Cancer Consortium, the Open Force Field Consortium, Cycle for Survival, a Louis V. Gerstner Young Investigator Award, and the Sloan Kettering Institute. A complete funding history for the Chodera lab can be found at http://choderalab.org/funding.

JM is a member of the Scientific Advisory Board of Cresset.

VSP currently sits on the boards of Apeel Sciences, BioAge, Devoted Health, Freenome, Genesis Therapeutics, Insitro, Inceptive Therapeutics, Nautilus Biotechnology, Nobell, Omada Health, Q.bio, SciFi Foods, and Scribe Therapeutics, and is on the Scientific Advisory Board of Schrodinger.

YW has limited financial interests in Flagship Pioneering, Inc. and its subsidiaries.